\documentclass{JHEP3}
\usepackage{amsmath}
\usepackage{epsfig}

\title{On QCD resummation with $k_t$ clustering}

\author{Yazid Delenda, Robert Appleby, Mrinal Dasgupta\\
        School of Physics and Astronomy, University of Manchester,\\
        Oxford road, Manchester M13 9PL, U.K.\\
        \email{yazid@hep.man.ac.uk},
        \email{Robert.Appleby@manchester.ac.uk},
        \email{Mrinal.Dasgupta@manchester.ac.uk}}

\author{Andrea Banfi\\
        Universit\`a degli studi di Milano Bicocca and INFN, Sezione
        di Milano, Italy.\\
        \email{Andrea.Banfi@mib.infn.it}}

\preprint{MAN/HEP/2006/27\\Bicocca-FT-06-16}

\abstract{We revisit the impact of the jet algorithm on predictions
of energy flow into gaps between hard jets, defined using the $k_t$
clustering procedure. The resulting prediction has two distinct
components: a primary emission piece that is related to independent
emission of soft gluons by the hard jets and a correlated emission
(non-global) piece known only in the large $N_c$ limit. We
analytically compute the dependence of the primary emission term on
the jet algorithm, which gives significantly more insight than our
previous numerical study of the same. We also point out that the
non-global component of the answer is reduced even more
significantly by the clustering than suggested previously in the
literature. Lastly we provide improved predictions for the latest
ZEUS photoproduction data, assessing the impact of our latest
findings.}


\begin{document}

\section{Introduction}

One of the most commonly studied QCD observables is the flow of
transverse energy ($E_t$) into gaps between jets in various QCD hard
processes. Since the $E_t$ flow away from jets is infrared and
collinear safe it is possible to make perturbative predictions for
the same, which can be compared to experimental data for a given
hard process. However since one is typically examining
configurations where $E_t$ is small compared to the hard scale $Q$
of the process (e.g. jet transverse momenta in hadronic collisions)
the perturbative predictions involve large logarithms in the ratio
$Q/E_t$. Resummation of logarithmically enhanced terms of the form
$\alpha_s^n \ln^{n} (Q/E_t)$ has proved a challenge that is still to
be fully met -- complete calculations are available only in the
large $N_c$ limit~\cite{DSNG1,DSNG2,BMS}. Studies of the $E_t$ flow
have in fact directly led to developments in the theoretical
understanding of QCD radiation and this process is still
ongoing~\cite{FKS}.

Another feature of the energy flow away from jets is its sensitivity
to non-perturbative effects. Thus one may expect significant $1/Q$
power corrections to energy flow distributions of a similar origin
to those extensively studied for various jet-shape
observables~\cite{DSreview}. Moreover the $E_t$ flow in hadronic
collisions is a standard observable used to develop an understanding
of the underlying event and to assess its role after accounting for
perturbatively calculable QCD radiation~\cite{MW,BKS}.

Given that $E_t$ flow studies potentially offer so much valuable
information on QCD over disparate scales, involving perturbative
parameters such as the strong coupling $\alpha_s$, QCD evolution,
coherence properties of QCD radiation and non-perturbative effects,
it is not surprising that they have been the subject of substantial
theoretical effort over the past few years.

In this paper we wish to focus on the aspect of resummed predictions
for the $E_t$ flow into gaps between jets. Perhaps the most
significant problem involved in making such predictions is the
non-global nature of the observable~\cite{DSNG1,DSNG2}. More
precisely in order to resum the leading single logarithms involved,
one has to address not just a veto on soft emissions coupled to the
underlying primary hard parton antennae (known as the primary
emission term), but additionally correlated emission or non-global
contributions, where a clump of energy-ordered soft gluons
coherently emit a still softer gluon into the gap region $\Omega$.
For this latter contribution the highly non-trivial colour structure
of the multi-gluon ``hedgehog" configuration has proved at present
too significant an obstacle to overcome. One thus has to resort to
the large $N_c$ limit to provide merely a leading-log estimate for
the away-from--jet $E_t$ flow. This  situation can be contrasted
with the case of event-shapes and Drell-Yan $q_T$ resummations which
have been pushed to next-to--leading and next-to--next-to--leading
logarithmic accuracy respectively. The impact of finite $N_c$
corrections in non-global observables is thus a factor in the
theoretical uncertainty involved in the corresponding resummed
predictions.

Given that the non-global component has a substantial quantitative
impact over a significant range of $E_t$ values for a given hard
scale $Q$ and that it is computable only in the large $N_c$
approximation, it is  clearly desirable to reduce the sensitivity of
a given observable to non-global logarithms. An important
observation in this regard was made in Ref.~\cite{AS1}: if one
employs the $k_t$ clustering algorithm~\cite{ktclus,ktclusinc} to
define the final state such that the energy flow into a gap between
jets is due to soft $k_t$-clustered mini-jets (rather than
individual hadrons), the non-global logarithms are significantly
reduced in magnitude\footnote{For recent progress on aspects of the
$k_t$ algorithm itself see Ref.~\cite{CacSal}.}. This observation
was exploited to study the case of $E_t$ flow in dijet
photoproduction where a result was provided for the primary emission
component of the $E_t$ distribution and the reduced non-global
component was modeled~\cite{AS2}.

However it has subsequently been found that $k_t$ clustering also
has a non-trivial impact on the primary emission component of the
result~\cite{BD05}. This was not taken into account in
Refs.~\cite{AS1,AS2} and also affects the ability to make resummed
predictions for a host of other jet observables such as azimuthal
correlation between jets $\Delta \phi_{jj}$. In fact the findings of
Ref.~\cite{BD05} are not just specific to the $k_t$ algorithm but
would also crop up in the case of jet observables defined using
iterative cone algorithms.

In the present paper we wish to shed more light on the resummation
of the primary or independent emission component of the result and
its dependence on the clustering algorithm. While the leading
${\mathcal{O}} \left (\alpha_s^2 \ln^2 (Q/E_t) \right)$ clustering
dependent behaviour was computed analytically in Ref.~\cite{BD05},
the full resummed result for the primary emission component was
computed only numerically in the case of a single hard emitting
dipole ($e^{+}e^{-} \to 2$ jets or DIS $1+1$ jets). Here while
sticking to a single hard dipole we shed more light on the structure
of the primary emission term and analytically compute it to an
accuracy that is sufficient for a wide range of phenomenological
applications.

The analytical insight and calculations we provide here will also
make the generalisation of the $k_t$-clustered primary emission
result to the case of several hard emitters (dijets produced in
photoproduction or hadron-hadron processes), involving a non-trivial
colour flow, relatively straightforward.

The above resummation of the primary component of the answer assumes
greater significance when we discuss our second observation: once an
error is corrected in the numerical code used for the purposes of
Refs.~\cite{AS1,AS2} the non-global component of the result is
reduced even more compared to the earlier estimate. With a very
small non-global component (which can be numerically computed in the
large $N_c$ limit) and a primary emission component that correctly
treats the dependence on the jet algorithm, one is better placed to
make more accurate resummed predictions than has been the case till
now. This is true not just for the $E_t$ flow but also as we
mentioned for a variety of jet observables for which there are
either no resummed predictions as yet, or only those employing jet
algorithms not directly used in experimental studies~\cite{KS}.

This paper is organised as follows. In the following section we
define the observable in question and revisit the issue of the
dependence of the primary and non-global pieces on the jet
clustering algorithm. Following this we demonstrate how the primary
or independent emission piece can be computed at all orders in
$\alpha_s$, accounting to sufficient accuracy for the effects of the
clustering algorithm. We explicitly describe the case of three and
four-gluon contributions to demonstrate the steps leading to our
all-order results. Following this we re-examine the non-global
component of the answer and find that this is significantly smaller
than earlier calculations of the same~\cite{AS1}. We put our
findings together to examine their impact on photoproduction data
from the ZEUS collaboration~\cite{ZEUS} and lastly point to the
conclusions one can draw and future extensions of our work.

\section{Resummation of the primary emissions}

Let us consider for simplicity the process $e^{+}e^{-} \to 2$ jets.
The calculations for processes involving a larger number of jets and
more complex jet topologies can be done along similar lines.

We wish to examine the $E_t$ flow in a region $\Omega$ which we
choose as a slice in rapidity\footnote{Since we are here dealing
with back-to--back jets we can define  the rapidity with respect to
the jet axis or equivalently, for our purposes, the thrust axis.} of
width $\Delta \eta$ which  we can centre on $\eta=0$. We then define
the gap transverse energy as:
\begin{equation}
\label{defn} E_t = \sum_{i \in \Omega}{E_{t,i}}\,,
\end{equation}
where the index $i$ refers to soft jets obtained after $k_t$
clustering of the final state. We shall concentrate on the
integrated $E_t$ cross-section which is defined as:
\begin{equation}
\Sigma(Q,Q_\Omega) = \frac{1}{\sigma}\int_0^{Q_\Omega}
\frac{d\sigma}{d E_t} d E_t\,,
\end{equation}
with $\sigma$ the total cross-section for $e^+e^-\rightarrow$
hadrons, with center-of-mass energy $Q$.

The single-logarithmic result for the above, without $k_t$
clustering (where the sum over $i$ in Eq.~\eqref{defn} refers to
hadrons in the gap rather than jet clusters), was computed in
Ref.~\cite{DSNG2} and can be expressed as:
\begin{equation}
\label{eq:signoclus} \Sigma(Q,Q_\Omega) = \Sigma_P (t)\,
S(t)\,,\qquad t = \frac{1}{2\pi} \int_{Q_\Omega}^{Q/2}
\frac{dk_t}{k_t}\, \alpha_s(k_t)\,.
\end{equation}
The above result contains a primary emission or ``Sudakov''
term\footnote{We use the term ``Sudakov'' in a loose sense since the
primary emission result leads to an exponential that is analogous to
a Sudakov form-factor.} $\Sigma_P(t)$ and a non-global term $S(t)$.

The primary emission piece is built up by considering only emissions
attached to the primary hard partons namely those emitted from the
hard initiating $q\bar{q}$ dipole in our example, while the
non-global term arises from coherent soft emission from a complex
ensemble of soft emitters alongside the hard initiating dipole. More
precisely we have:
\begin{equation}
\label{eq:sudakov} \Sigma_P(t) = e^{-4 C_F t\Delta \eta }\,,
\end{equation}
which is the result of resumming uncancelled $k_t$-ordered
virtual-emission contributions, in the gap region. The non-global
component, as we stated before, is computed numerically in the large
$N_c$ limit.

Next we turn to the $k_t$-clustered case. The result stated in
Ref.~\cite{AS1} assumes that the primary or Sudakov piece is left
unchanged by clustering since it appears to be the exponentiation of
a single gluon emitted inside the gap. The non-global piece is
recomputed numerically implementing clustering~\cite{AS1}. As
already shown in Ref.~\cite{BD05} however, the assumption regarding
the primary emission piece being unaffected is in fact untrue and
this too needs to be recomputed in the presence of clustering. The
corrections to the primary emission term first appear while
considering two gluons emitted by the hard $q \bar{q}$ dipole and
persist at all orders. Below we provide a reminder of the two-gluon
case discussed in Ref.~\cite{BD05} and subsequently consider
explicitly the three and four-gluon emission cases before writing
down the result to all orders as a function of the radius parameter
$R$.

\subsection{Two-gluon emission}

In order to examine the role of the $k_t$ algorithm we point out
that in our case ($k_t$-ordered soft limit) one can start the
clustering procedure with the lowest transverse-energy parton or
equivalently the softest parton. One examines the ``distances'' of
this particle, $i$, from its neighbours, defined by $d_{ij} =
E_{t,i}^2 \left(\left(\Delta \eta_{ij} \right)^2 + \left( \Delta
\phi_{ij}\right)^2\right)$, where $E_{t,i}$ is the transverse energy
of the softest parton. If the smallest of these distances is less
than $E_{t,i}^2 R^2$, particle $i$ is recombined or clustered into
its nearest neighbour  and the algorithm is iterated. On the other
hand if all $d_{ij}$ are greater than $E_{t,i}^2 R^2$, $i$ is
counted as a jet and removed from the process of further clustering.
The process continues until the entire final-state is made up of
jets. Also in the limit of strong energy-ordering, which is
sufficient to obtain the leading-logarithms we are concerned with
here, the recombination of a softer particle with a harder one gives
a jet that is aligned along the harder particle.

The dependence of  the primary emission term on the jet clustering
algorithm starts naturally enough from the two-gluon level. While
the Sudakov result $\exp{\left(-4 C_F t\Delta \eta\right)}$ comes
about due to assuming real-virtual cancellations such that one is
left with only virtual emissions with $k_t \geq Q_{\Omega}$ in the
gap region (for the integrated distribution), $k_t$ clustering
spoils this assumed cancellation.

Specifically let us take two real gluons $k_1$ and $k_2$ that are
ordered in energy ($\omega_1 \gg \omega_2$). We consider as in
Ref.~\cite{BD05} the region where the softer gluon $k_2$ is in the
gap whilst the harder $k_1$ is outside. Additionally we take the
case that the gluons are clustered by the jet algorithm which
happens when $\left(\Delta \eta \right)^2 + \left( \Delta \phi
\right)^2 \leq R^2$ with $\Delta\eta = \eta_2-\eta_1$ and similarly
for $\Delta \phi$, which condition we shall denote with the symbol
$\theta_{21}$. Since $k_2$ is clustered to $k_1$ it gets pulled
outside the gap, the recombined jet being essentially along $k_1$.
Thus in this region the double real-emission term does not
contribute to the gap energy \emph{differential} distribution
$d\sigma/dE_t$. Now let $k_1$ be a virtual gluon. In this case it
cannot cluster $k_2$ out of the gap and we do get a contribution to
the gap energy differential distribution. Thus a real-virtual
cancellation which occurs for the unclustered case fails here and
the mismatch for the integrated quantity $\Sigma(t)$, amounts to:
\begin{equation}
\label{eq:twog} C_2^{p} = \frac{(-4 C_F t)^2}{2!}
\int_{k_1\notin\Omega} d\eta_1 \frac{d \phi_1}{2 \pi} \int_{ k_2 \in
\Omega} d \eta_2 \frac {d \phi_2}{2 \pi} \theta_{21} = \frac{(-4C_F
t)^2}{2!} \frac{2}{3 \pi} R^3\,,
\end{equation}
where we reported above the result computed for $R \leq \Delta
\eta$, in Ref.~\cite{BD05}. Here we introduced the primary emission
term $C_n^{p}$ that corrects the Sudakov result at
$\mathcal{O}(\alpha_s^n)$ due to the clustering requirement.

The fact that the result scales as the third power of the jet radius
parameter is interesting in that by choosing a sufficiently small
value of $R$ one may hope to virtually eliminate this piece and thus
the identification of the primary result with the Sudakov exponent
would be at least numerically accurate. However the non-global term
would then be significant which defeats the main use of clustering.
If one chooses to minimise the non-global component by choosing e.g.
$R=1$, then one must examine the primary emission terms in higher
orders in order to estimate their role. To this end we start by
looking at the three and four-gluon cases below.

\subsection{Three-gluon emission}

Consider the emission of three energy-ordered gluons $k_1$, $k_2$
and $k_3$ with $\omega_3\ll \omega_2 \ll \omega_1$, off the primary
$q\bar{q}$ dipole, and employing the inclusive $k_t$ clustering
algorithm~\cite{ktclus,ktclusinc} as explained previously.

We consider all the various cases that arise when the gluons (which
could be real or virtual) are in the gap region or outside. We also
consider all the configurations in which the gluons are affected by
the clustering algorithm. We then look for all contributions where a
real-virtual mismatch appears due to clustering, that is not
included in the exponential Sudakov term. The Sudakov itself is
built up by integrating just virtual gluons in the gap, above the
scale $Q_\Omega$. The corrections to this are summarised in
table~\ref{tab:cont}. \TABLE[ht]{
\begin{tabular}{|c|c|c||c|c|c|c|c|c|}
\hline $\theta_{32}$ & $\theta_{31}$ & $\theta_{21}$ &
$k_3\in\Omega$ & $k_2 \in\Omega$ & $k_1\in\Omega$ & $k_3\,,\,k_2
\in\Omega$ & $k_3\,,\,k_1 \in\Omega$ & $k_2\,,\,k_1 \in\Omega$
\\
\hline\hline
0 & 0 & 0 & 0 & 0 & 0 & 0 & 0 & 0\\
\hline
1 & 0 & 0 & 0 & 0 & 0 & 0 &$W$& 0\\
\hline
0 & 1 & 0 & 0 & 0 & 0 &$W$& 0 & 0\\
\hline
0 & 0 & 1 & 0 & 0 & 0 &$W$& 0 & 0\\
\hline
1 & 1 & 0 &$W$& 0 & 0 &$W$&$W$& 0\\
\hline
1 & 0 & 1 & 0 & 0 & 0 & 0 &$W$& 0\\
\hline
0 & 1 & 1 & 0 & 0 & 0 &$W$& 0 & 0\\
\hline
1 & 1 & 1 &$W$& 0 & 0 &$W$&$W$& 0\\
\hline
\end{tabular}
\caption{\label{tab:cont}Contributions of different configurations
of particles to $\Sigma_P(t)\,$ at $\mathcal{O}(\alpha_s^3)$. We
define $\theta_{ij} = \theta\left( R^2-(\eta_{i}-\eta_j)^2 -
(\phi_{i} - \phi_j)^2\right)$, e.g. $\theta_{13}=1$ means
$(\eta_{1}-\eta_3)^2+(\phi_{1}-\phi_3)^2\leq R^2$. We also define
$W=(-4C_Ft)^3 /3!$, so the entries ``$W$'' indicate a
miscancellation which leads to a single-log correction to the
Sudakov result, while the entries ``0'' indicate a complete
real-virtual cancellation. We have discarded the case where all
particles are in the gap since such configurations are already
included in the exponential Sudakov result.}}

In order to obtain the various entries of the table one just looks
at the angular configuration in question, draws all possible real
and virtual contributions and looks for a mismatch between them
generated by the action of clustering. We translate
table~\ref{tab:cont} to:
\begin{eqnarray}\label{eq:cont}
C_3^{p} & = & \frac{1}{3!}(-4C_F t)^3\times \nonumber\\
& & \times \bigg\{ \int_{k_1\notin\Omega} d\eta_1
\frac{d\phi_1}{2\pi} \int_{k_2\notin\Omega} d\eta_2
\frac{d\phi_2}{2\pi} \int_{k_3\in\Omega} d\eta_3\,
\theta_{32}\, \theta_{31} + \nonumber\\
& & + \int_{k_1\notin\Omega} d\eta_1 \frac{d\phi_1}{2\pi}
\int_{k_2\in\Omega} d\eta_2 \frac{d\phi_2}{2\pi} \int_{k_3\in\Omega}
d\eta_3 \left[\theta_{31}+(1-\theta_{31})(1-\theta_{32})
\theta_{21}\right] + \nonumber\\
& & + \int_{k_1\in\Omega} d\eta_1 \frac{d\phi_1}{2\pi}
\int_{k_2\notin\Omega} d\eta_2 \frac{d\phi_2}{2\pi}
\int_{k_3\in\Omega} d\eta_3\, \theta_{32}\bigg\}\,,
\end{eqnarray}
where we used the freedom to set $\phi_3=0$. We identify three equal
contributions consisting of the integrals in which there is only one
theta function constraining only two particles: the last integral
over $\theta_{32}$, the integral over $\theta_{31}$ and that over
$\theta_{21}$ in the third line. The set of configurations
$\theta_{32}$, $\theta_{31}$ and $\theta_{21}$  is just the set of
constraints on all possible pairs of gluons, and in fact we can
generalise the factor 3 to the case of any number $n$ of gluons by
$n(n-1)/2$, which will enable us to resum $R^3$ terms. We shall
return to this observation later. The integrals of the above type
reduce essentially to the clustered two-gluon case as calculated in
Eq.~\eqref{eq:twog}, and the integral over the third
``unconstrained'' gluon is just $\Delta\eta$.

Explicitly we write Eq.~\eqref{eq:cont} as:
\begin{eqnarray}
\label{eq:cont2}
C_3^{p} & = & \frac{1}{3!} (-4C_F t)^3 \times \nonumber\\
& & \times\Bigg\{ \int_{k_1\notin\Omega} d\eta_1
\frac{d\phi_1}{2\pi} \int_{k_2\notin\Omega} d\eta_2
\frac{d\phi_2}{2\pi} \int_{k_3\in\Omega} d\eta_3\,
\theta_{32}\,\theta_{31}+\nonumber\\
& & + \int_{k_1\notin\Omega} d\eta_1 \frac{d\phi_1}{2\pi}
\int_{k_2\in\Omega} d\eta_2 \frac{d\phi_2}{2\pi} \int_{k_3\in\Omega}
d\eta_3 \left[\theta_{31}\theta_{32}-\theta_{31}-\theta_{32}
\right] \theta_{21}+ \nonumber\\
&& + 3\times \int_{k_1\in\Omega} d\eta_1 \frac{d\phi_1}{2\pi}
\int_{k_2\notin\Omega} d\eta_2 \frac{d\phi_2}{2\pi}
\int_{k_3\in\Omega} d\eta_3\,\theta_{32}\Bigg\}\,.
\end{eqnarray}
Computing the various integrals above (for simplicity we take $R
\leq \Delta \eta/2$, which is sufficient for our phenomenological
purposes) one obtains:
\begin{multline}
C_3^{p} = \frac{1}{3!} (-4C_F t)^3 \times\\
\times \left\{ \left( \frac{\pi}{3}-\frac{32}{45} \right)
\frac{R^5}{\pi^2} + f \frac{R^5}{\pi^2}- \left(\frac{\pi}{3} -
\frac{32}{45} \right) \frac{R^5}{\pi^2} - \frac{32}{45}
\frac{R^5}{\pi^2} + 3 \times \frac{2}{3\pi} \Delta\eta \, R^3
\right\},
\end{multline}
with $f \simeq 0.2807$ and we have written the results in the same
order as the five integrals that arise from the various terms in
Eq.~\eqref{eq:cont2}. Hence:
\begin{equation}
C_3^{p}=\frac{1}{3!}(-4C_F t)^3 \left\{3\times \frac{2}{3\pi}
\Delta\eta \, R^3 + f_2\, R^5 \right\},
\end{equation}
where $f_2 \simeq -0.04360$. We note the appearance of an $R^5$ term
which, as we shall presently see, persists at higher orders. This
term is related to a clustering constraint on \emph{three} gluons at
a time via the product of step functions $\theta_{32}\, \theta_{21}
(\theta_{31}-1)\,$ with $k_2,\,k_3\in\Omega$ and $k_1\notin\Omega$.

Next we look at the emission of four soft, real or virtual
energy-ordered gluons. This will help us move to a generalisation
with any number of gluons.

\subsection{Four-gluon case and beyond}

Now we take the case of four-gluon emission and identify the
patterns that appear at all orders. A table corresponding to table
\ref{tab:cont} is too lengthy to present here. The result can
however be expressed in an equation similar to that for the
three-gluon case. We have:
\begin{eqnarray}
C_4^{p} & = & \frac{1}{4!} (-4C_Ft)^4 \times \nonumber\\
& & \times \bigg \{ \int_{1\,\mathrm{in}} \int_{2\,\mathrm{in}}
\int_{3\,\mathrm{out}} \int_{4\,\mathrm{in}} \theta_{43}+\nonumber\\
& & + \int_{1\,\mathrm{in}} \int_{2\,\mathrm{out}}
\int_{3\,\mathrm{in}} \int_{4\,\mathrm{in}}
\left[\theta_{42}+\theta_{32} (1-\theta_{43})
(1-\theta_{42})\right] + \nonumber\\
& & + \int_{1\,\mathrm{out}} \int_{2\,\mathrm{in}}
\int_{3\,\mathrm{in}} \int_{4\,\mathrm{in}} \left\{ \theta_{41} +
\theta_{-41} \left[ \theta_{31} \, \theta_{-43} + \theta_{43} \,
\theta_{21} \, \theta_{-42}+ \theta_{21} \, \theta_{-42}\,
\theta_{-43}\,\theta_{-31} \theta_{-32} \right] \right\} +
\nonumber\\ & & + \int_{1\,\mathrm{in}} \int_{2\,\mathrm{out}}
\int_{3\,\mathrm{out}} \int_{4\,\mathrm{in}}
\theta_{42}\, \theta_{43} + \nonumber\\
& & + \int_{1\,\mathrm{out}} \int_{2\,\mathrm{in}}
\int_{3\,\mathrm{out}} \int_{4\,\mathrm{in}} \theta_{43} \left[
\theta_{41}+ \theta_{-41} \, \theta_{-42}\, \theta_{21} \right] +
\nonumber\\ & & + \int_{1\,\mathrm{out}} \int_{2\,\mathrm{out}}
\int_{3\,\mathrm{in}} \int_{4\,\mathrm{in}} \theta_{41} \,
\theta_{42} + \theta_{41}\, \theta_{-42} \, \theta_{-43}\,
\theta_{32} + \theta_{-41}\, \theta_{-43} \, \theta_{31}
\left[\theta_{42} +
\theta_{-42}\, \theta_{32} \right] + \nonumber\\
& & + \int_{1\,\mathrm{out}} \int_{2\,\mathrm{out}}
\int_{3\,\mathrm{out}} \int_{4\,\mathrm{in}} \theta_{41}\,
\theta_{42}\, \theta_{43} \bigg\}\,,
\end{eqnarray}
where $\theta_{-ij}=1-\theta_{ij}$ and ``in'' or ``out'' pertains to
whether the gluon is inside the gap region or out. For brevity we
did not write the differential phase-space factor for each gluon
which is as always $d\eta\,d\phi/(2\pi)$. We identify six $R^3$
terms exactly of the same kind as computed before and similarly four
$R^5$ terms. Explicitly we have:
\begin{eqnarray}
C_4^{p} & = & \frac{1}{4!} (-4C_Ft)^4 \times \nonumber\\
& & \times \bigg\{ 6 \times \int_{1\,\mathrm{in}}
\int_{2\,\mathrm{in}} \int_{3\,\mathrm{out}}
\int_{4\,\mathrm{in}} \theta_{43}+\nonumber\\
& & + 4 \times \left( \int_{1\,\mathrm{in}} \int_{2\,\mathrm{out}}
\int_{3\,\mathrm{out}} \int_{4\,\mathrm{in}} \theta_{42}\,
\theta_{43} + \int_{1\,\mathrm{in}} \int_{2\,\mathrm{out}}
\int_{3\,\mathrm{in}} \int_{4\,\mathrm{in}} \theta_{32}
\left[\theta_{43} \, \theta_{42} - \theta_{43} - \theta_{42}
\right] \right) + \nonumber\\
& & + 3 \times \int_{1\,\mathrm{out}} \int_{2\,\mathrm{in}}
\int_{3\,\mathrm{out}} \int_{4\,\mathrm{in}} \theta_{21}\,
\theta_{43}
\left[1-\theta_{41}-\theta_{42}+\theta_{41}\,\theta_{42}\right]
+\nonumber\\
& & + \int_{1\,\mathrm{out}} \int_{2\,\mathrm{in}}
\int_{3\,\mathrm{in}} \int_{4\,\mathrm{in}} \theta_{21}
\big[\theta_{42}\,\theta_{43}-\theta_{42}-\theta_{43}-\theta_{41}\,
\theta_{-42}\, \theta_{-43}\big] \big[\theta_{31} \,
\theta_{32}-\theta_{31} -\theta_{32} \big] + \nonumber\\
& & + \int_{1\,\mathrm{out}} \int_{2\,\mathrm{out}}
\int_{3\,\mathrm{in}} \int_{4\,\mathrm{in}} \theta_{32}\,
\theta_{31} \left[\theta_{41}(1-\theta_{43})(\theta_{42}-2)
-\theta_{43}\right] +\nonumber\\
&&+\int_{1\,\mathrm{out}}\int_{2\,\mathrm{out}}\int_{3\,\mathrm{out}}
\int_{4\,\mathrm{in}}\theta_{41}\,\theta_{42}\,\theta_{43}
\bigg\}\,.
\end{eqnarray}
We discuss below each set of integrals, generalise the result to the
case of $n$ emitted gluons and then resum all orders.

\begin{itemize}
\item The integral:
\begin{equation}
\frac{1}{4!} (-4C_Ft)^4\,6 \times \int_{1\,\mathrm{in}}
\int_{2\,\mathrm{in}} \int_{3\,\mathrm{out}} \int_{4\,\mathrm{in}}
\theta_{43}\,.
\end{equation}
\end{itemize}
The integrals over particles 1 and 2 give $\left (\Delta\eta
\right)^2$. The remaining integrals reduce to the result computed
for the two-gluon case, i.e. the $R^3$ term, multiplied by a factor
of 6 accounting for the number of pairs of gluons $n(n-1)/2$, for
$n=4$. Explicitly we have for this term:
\begin{equation}
\frac{1}{4!} (-4C_Ft)^4 \frac{4\times
3}{2}\Delta\eta^{4-2}\frac{2}{3\pi}R^3\,.
\end{equation}
For $n$ emitted gluons the $R^3$ term, which is always related to
the clustering of two gluons, is given by:
\begin{equation}
\frac{1}{n!} \frac{n(n-1)}{2} (-4C_Ft\Delta\eta)^n \Delta\eta^{-2}
\frac{2}{3\pi} R^3\,, \quad n\geq 2\,.
\end{equation}
Hence to all orders one can sum the above to obtain:
\begin{equation}
e^{-4C_Ft\Delta\eta}\frac{(-4C_Ft)^2}{2}\frac{2}{3\pi}R^3\,.
\end{equation}
\begin{itemize}
\item The integrals:
\begin{multline}
\frac{1}{4!}(-4C_Ft)^4\,4\times \bigg( \int_{1\,\mathrm{in}}
\int_{2\,\mathrm{out}} \int_{3\,\mathrm{out}}
\int_{4\,\mathrm{in}} \theta_{42}\,\theta_{43}+\\
+\int_{1\,\mathrm{in}} \int_{2\,\mathrm{out}}\int_{3\,\mathrm{in}}
\int_{4\,\mathrm{in}}\theta_{32} \left[\theta_{43}\, \theta_{42}-
\theta_{43} - \theta_{42} \right] \bigg)\,.
\end{multline}
\end{itemize}
The integral over particle 1 gives $\Delta\eta$, while the rest of
the integrals reduce to the ones calculated earlier which gave the
$R^5$ result, accompanied with a factor of $4$ standing for the
number of triplet combinations formed by four gluons. For $n$
emitted gluons this factor is $n(n-1)(n-2)/3!$. Explicitly we have
for this case:
\begin{equation}
\frac{1}{4!} (-4C_Ft)^4\frac{4\times 3\times 2}
{6}\Delta\eta^{4-3}f_2\,R^5\,.
\end{equation}
At the $n^{\mathrm{th}}$ order we obtain:
\begin{equation}
\frac{1}{n!} (-4C_Ft\Delta\eta)^n \frac{n(n-1)(n-2)} {6}
\Delta\eta^{-3} f_2\,R^5\,, \quad n\geq 3\,.
\end{equation}
Summing all orders we get:
\begin{equation}
e^{-4C_Ft\Delta\eta} \frac{(-4C_Ft)^3}{6} f_2\, R^5\,.
\end{equation}
\begin{itemize}
\item The integral:
\begin{equation}
\frac{1}{4!}(-4C_Ft)^4\,3\times \int_{1\,\mathrm{out}}
\int_{2\,\mathrm{in}} \int_{3\,\mathrm{out}} \int_{4\,\mathrm{in}}
\theta_{21}\, \theta_{43}\,.
\end{equation}
\end{itemize}
This integral can be factored into two separate integrals involving
the constraint on $k_1$ and $k_2$ and over $k_3$ and $k_4$
respectively. Each of these reduces to the $R^3$ result obtained in
the two-gluon case. Thus we get:
\begin{equation}
\frac{1}{4!} (-4C_Ft)^4\, 3\times\left(\frac{2}{3\pi}\right)^2R^6\,.
\end{equation}
At $n^{\mathrm{th}}$ order this becomes:
\begin{equation}
\frac{1}{n!} \frac{n(n-1)(n-2)(n-3)}{8}(-4C_F t\Delta\eta)^n
\Delta\eta^{-4} \left(\frac{2}{3\pi}\right)^2R^6\,,\quad n\geq 4\,,
\end{equation}
which can be resummed to:
\begin{equation}
e^{-4C_Ft\Delta\eta}\frac{(-4C_Ft)^4}{8}
\left(\frac{2}{3\pi}\right)^2R^6.
\end{equation}
The factor 3 (and generally $n(n-1)(n-2)(n-3)/8$) is the number of
configurations formed by four (and generally $n$) gluons such that
we have two pairs of gluons each is formed by an out-of-gap gluon
connected to a softer in-gap one.
\begin{itemize}
\item The remaining integrals
\end{itemize}
These integrals give at most an  $\mathcal{O}(R^7)$ term because
they constrain all the four gluons at once. In fact for gap sizes
$\Delta\eta \geq 3R$, these integrals go purely as $R^7$ with no
dependence on $\Delta \eta$. Since here however we wish to use the
condition $\Delta \eta \geq 2R$, which allows us to make use of the
whole range of HERA data, these integrals do not depend purely on
$R$ but are a function of $R$ and $\Delta \eta$ which have an upper
bound of order $R^7$. This can be seen by noting that there are
three azimuthal integrations that each produce a function which has
a maximum value proportional to $R$, so the result of integrating
over all azimuthal variables is a factor that is bounded from above
by $R^3$. Similarly there are four rapidity integrations with a
clustering constraint on all four gluons implying that they can
produce an $R^4$ term at most. In general the result at
$n^\mathrm{th}$ order of constraining $n$ gluons at once, is bounded
from above by a factor of order $R^{2n-1}$.

We can write the result for all these as $(-4C_Ft)^4/4!\,
y(R,\Delta\eta)$, and resum such terms to all orders (in the same
manner as before) to:
\begin{equation}
e^{-4C_Ft\Delta\eta}\frac{(-4C_Ft)^4}{4!} y(R,\Delta \eta)\,,
\end{equation}
where $y(R,\Delta \eta)$ is at most ${\mathcal{O}}(R^7)$. We do not
calculate these terms (though it is possible to do so) since the
accuracy we achieve by retaining the $R^3$, $R^5$ and $R^6$ terms,
we have already computed, is sufficient as we shall show.

The five-gluon case is too lengthy to analyse here. The same
patterns as pointed out above persist here but new terms that are at
most ${\mathcal{O}}(R^{9})$ appear when all five gluons are
constrained. There is also an $R^8$ term, coming from the
combination of $R^3$ and $R^5$ terms in the same manner that the
$R^6$ term arose as a combination of two $R^3$ terms.

\section{All-orders result}

From the above observations we can assemble an all-orders result to
$R^6$ accuracy, where we shall consider $R$ to be at most equal to
unity. The final result for primary emissions alone and including
the usual Sudakov logarithms (for $\Delta \eta \geq 2R$) is:
\begin{multline}\label{eq:result}
\Sigma_{P}(t) = e^{-4C_Ft\Delta\eta} \times\\ \times \left( 1+
(-4C_Ft)^2 \frac{1}{3\pi} R^3 + (-4C_Ft)^3 \frac{f_2} {6} R^5 +
(-4C_Ft)^4 \frac{1}{18\pi^2} R^6 + \frac{(-4C_Ft)^4}{4!}
\mathcal{O}(R^7) \right).
\end{multline}

Formally one may wish to extend this accuracy by computing a few
more terms such as those integrals that directly give or are bounded
by an $R^7$ behaviour and this is possible though cumbersome. It
should also be unnecessary from a practical viewpoint, especially
keeping in mind that $R=0.7$ is a preferable value to $R=1$, in the
important case of hadron collisions\footnote{This is because the
underlying event will contaminate jets less if one chooses a smaller
$R$.} and the fact that even at $R=1$ the $R^3$ term significantly
dominates the result over the range of $t$ values of
phenomenological interest, as we shall see below.

We further note that if one keeps track of all the terms that come
about as a combination of $R^3$ and $R^5$ terms in all possible ways
at all orders, one ends up with the following form for
Eq.~\eqref{eq:result}:
\begin{equation}
\Sigma_{P}(t)= e^{-4C_Ft\Delta\eta} \exp
\left(\frac{(-4C_Ft)^2}{2!}\frac{2}{3\pi}R^3 +
\frac{(-4C_Ft)^3}{3!}f_2\,
R^5+\frac{(-4C_Ft)^4}{4!}\mathcal{O}(R^7)\right),
\end{equation}
the expansion of which agrees with Eq.~\eqref{eq:result}. In the
above by ${\mathcal{O}}(R^7)$ we mean terms that, while they may
depend on $\Delta \eta$, are at most as significant as an $R^7$
term. We also mention that in the formal limit $\Delta \eta \to
\infty$, there is no dependence of the clustering terms on $\Delta
\eta$ and they are a pure power series in $R$. The limit of an
infinite gap appears in calculations where the region considered
includes one of the hard emitting partons. An example of such cases
(which have a leading double-logarithmic behaviour) is once again
the quantity $\Delta \phi_{jj}$ between jets in e.g. DIS or hadron
collisions.

\FIGURE[ht]{\epsfig{file=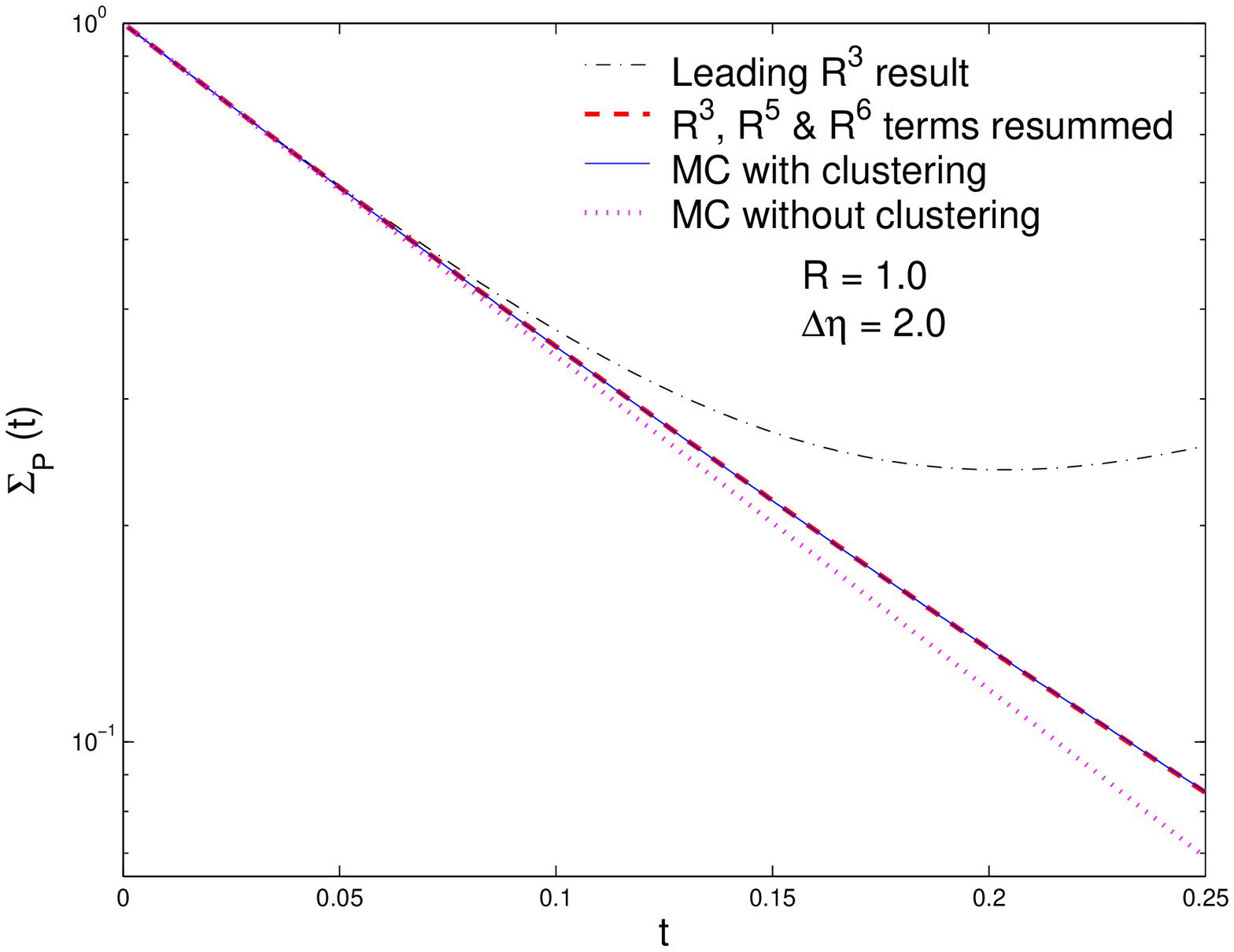,width=0.7\textwidth}
\caption{\label{fig:results}Comparison of the analytical results to
a numerical Monte Carlo estimate.}}

Fig.~\ref{fig:results} represents a comparison between the leading
$R^3$ result (i.e. the pure fixed-order result of Ref.~\cite{BD05}
combined with the resummed Sudakov exponent), the resummed $R^3$,
$R^5$ and $R^6$ result (Eq.~\eqref{eq:result}) and a numerical Monte
Carlo estimate with and without clustering. The Monte Carlo program
in question is essentially that described in Ref.~\cite{DSNG1}, with
the modification of $k_t$ clustering where we computed just
emissions off the primary dipole ``switching off'' the non-global
correlated emission.

We note that the resummed analytical form~\eqref{eq:result} is in
excellent agreement with the numerical result which contains the
full $R$ dependence. We have tested this agreement with a range of
values of $R$. We take this agreement as indicating that uncomputed
$R^7$ and higher terms can safely be ignored even at $R=1$ and even
more so at fractional values of $R$, e.g. $R = 0.7$. To provide an
idea about the relative role of terms at different powers of $R$ in
Eq.~\eqref{eq:result} we note that for $R=1$ and $t = 0.25$ the
resummed $R^3$ term increases the Sudakov result $\exp\left(-4 C_Ft
\Delta \eta \right)$ by $19 \%$, the $R^5$ term represents a further
increase of $1.5 \%$ to the result after inclusion of the resummed
$R^3$ term and the $R^6$ term has a similar effect on the result
obtained after including up to $R^5$ terms.

Next we comment on the size of the non-global component at different
values of $R$.

\section{Revisiting the non-global contribution}

We have seen above how the primary emission piece is dependent on
the jet clustering algorithm. It was already noted in
Ref.~\cite{AS1} that the non-global contribution is significantly
reduced by clustering. Here we wish to point out that after
correction of an oversight in the code used there, the non-global
component is even more significantly reduced than previously stated
in the literature. Indeed for $R=1$ and the illustrative value of
$t=0.15$, which corresponds to gap energy $Q_\Omega=1$ GeV for a
hard scale $Q =100$ GeV, the non-global logarithms are merely a $5
\%$ effect as opposed to the $20 \%$ reported previously~\cite{AS1}
and the over $65 \%$ effect in the unclustered case.

\FIGURE[ht]{\epsfig{file=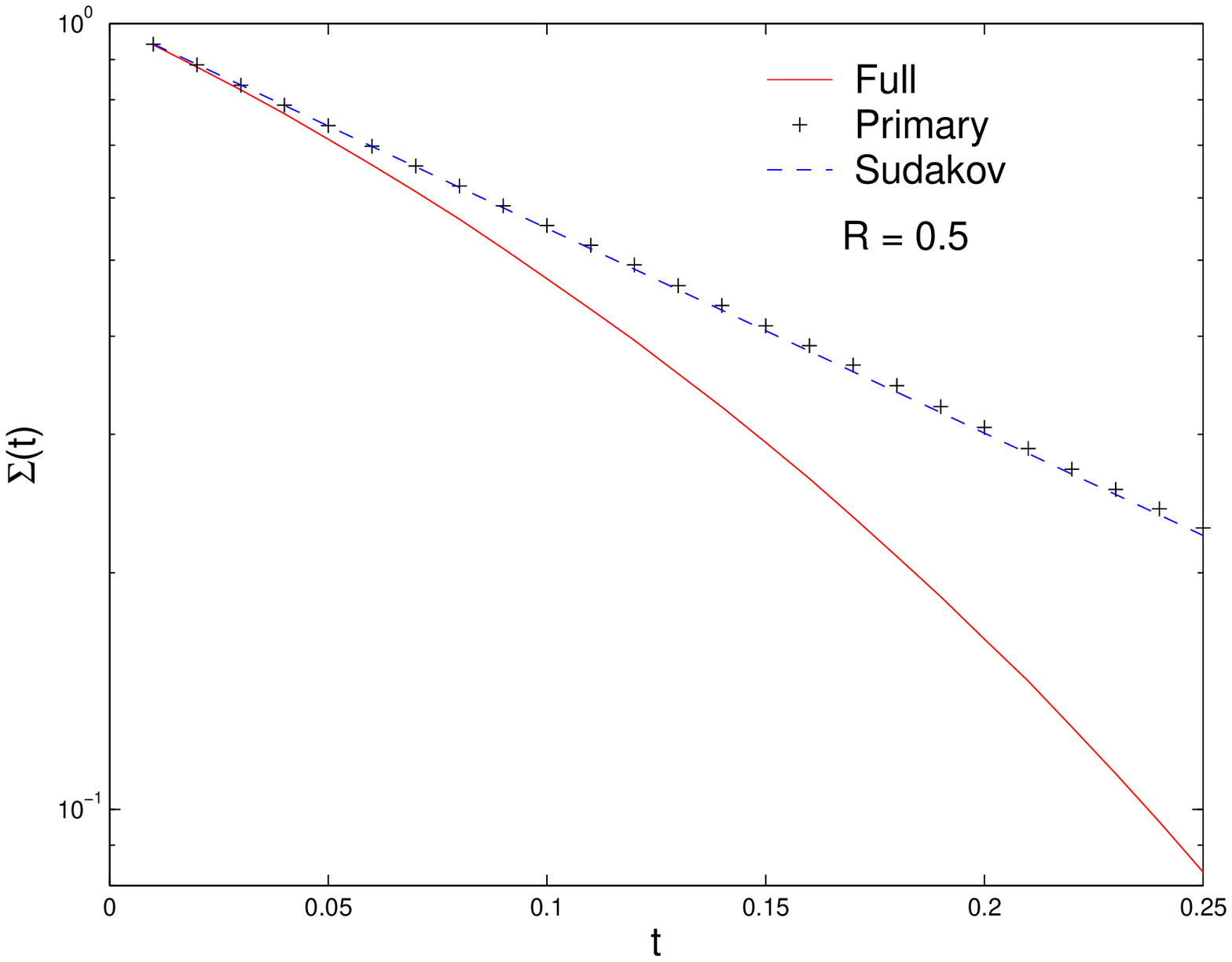, width = 0.55\textwidth}
\epsfig{file=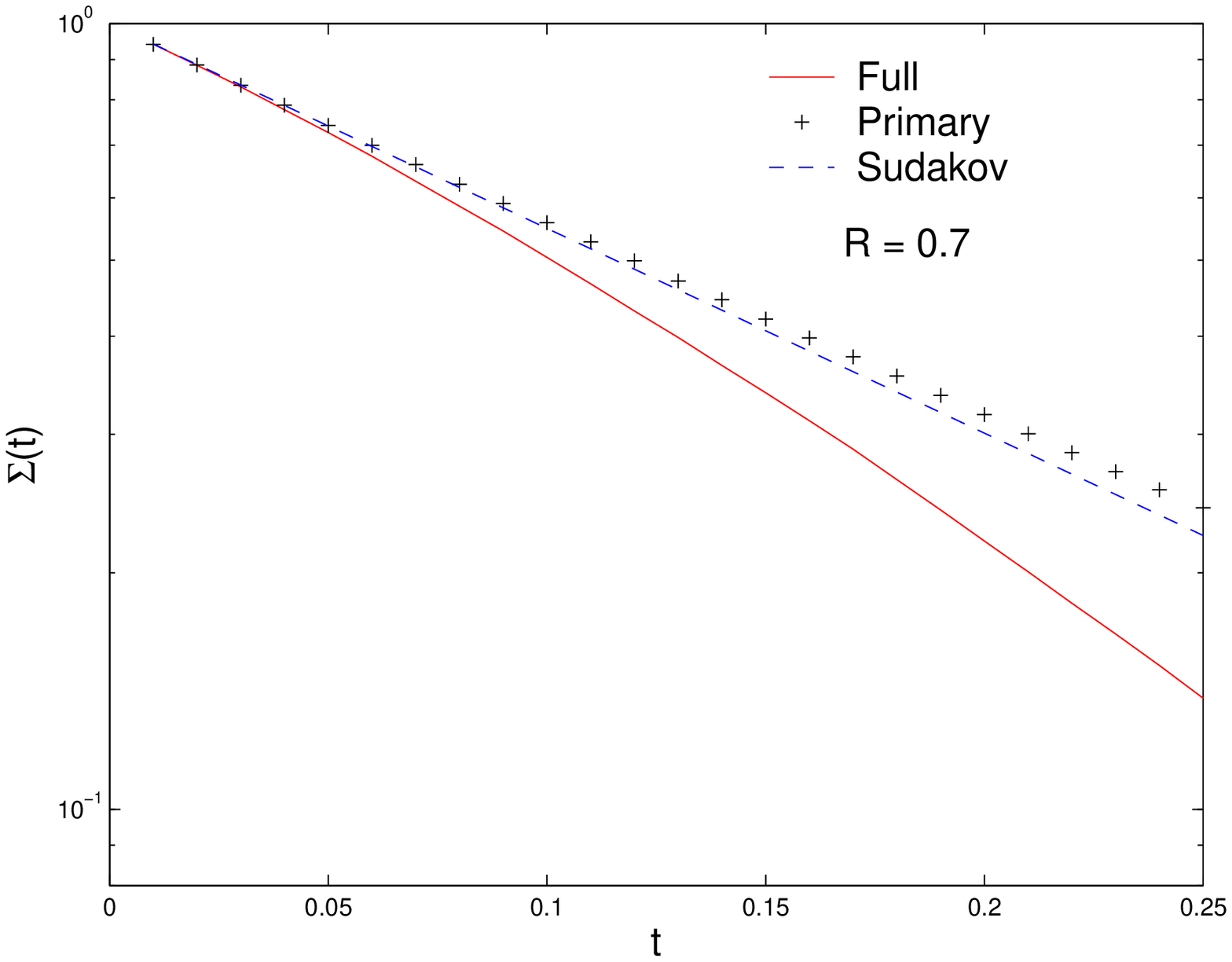, width = 0.55\textwidth}
\epsfig{file=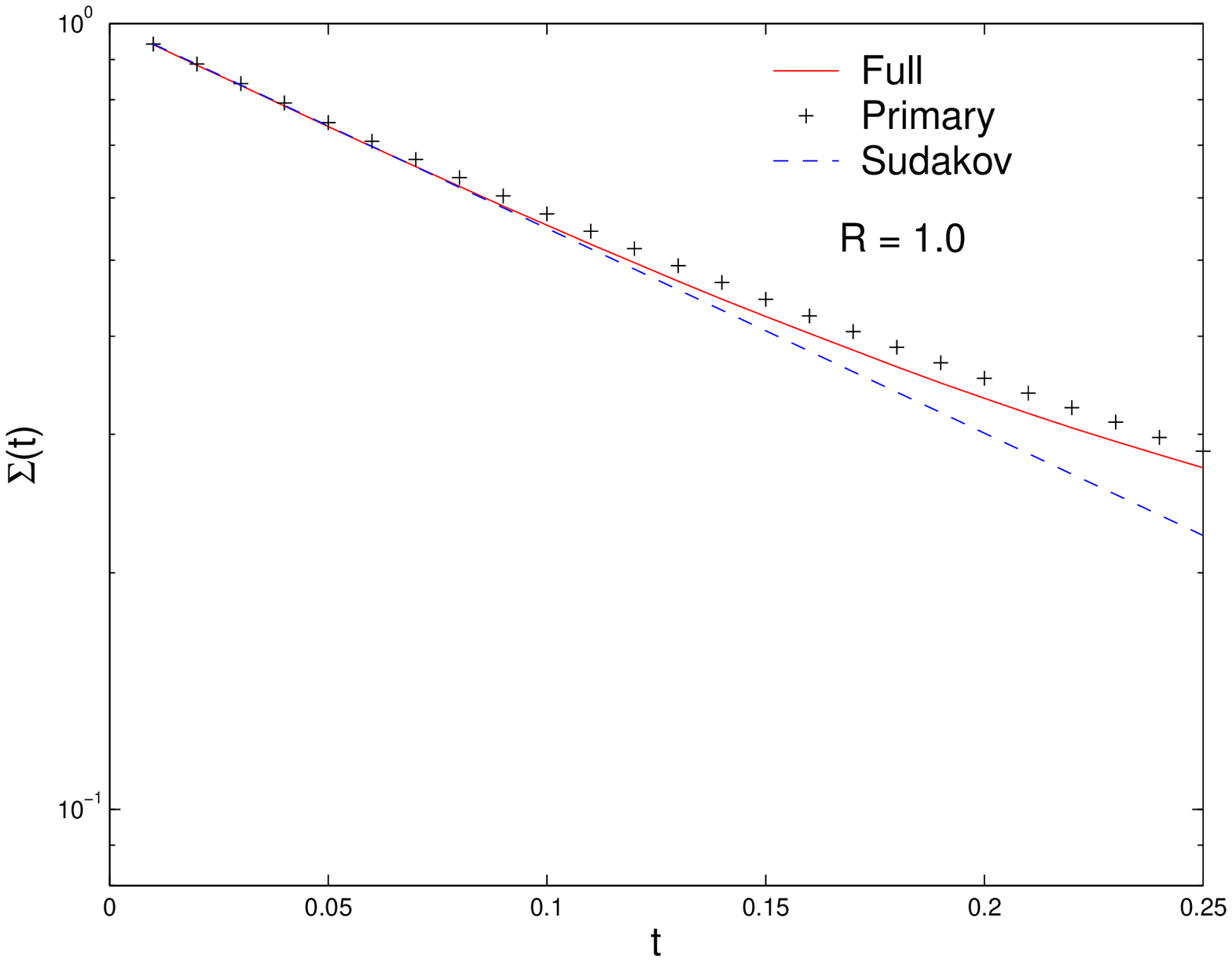, width =0.55\textwidth}
\caption{\label{fig:ngs}Comparison of the Sudakov result, the
correct primary result and the full result including non-global
logarithms, for different values of $R$ and with $\Delta \eta=1$.
All quantities are shown in the large $N_c$ limit for ease of
comparison.}}

In Fig.~\ref{fig:ngs} we plot the curves for the primary and full
results (in the large $N_c$ limit) for the integrated quantity
$\Sigma(t)$ as a function of $t$ defined earlier. We note that for
$R=0.5$ the primary result is essentially identical to the Sudakov
result. The non-global contribution (which is the ratio of the full
and primary curves) is however still quite significant. Neglecting
it leads to an overestimate of $40 \%$ for $t =0.15$. Increasing the
jet radius in a bid to lower the non-global component we note that
for $R=0.7$ the impact of the non-global component is now just over
$20 \%$ while the difference between the full primary result and the
Sudakov result is small (less than $5 \%$). The situation for $R=1$
is a bit different. Here it is the non-global logarithms that are
only a $5\%$ effect (compared to the $20 \%$ claimed
earlier~\cite{AS1}) while the full primary result is bigger than the
Sudakov term by around $11 \%$.

The value $R=1$ is in fact the one used in the HERA analyses of
gaps-between--jets in photoproduction. It is now clear that such
analyses will have a very small non-global component and a moderate
effect on primary emissions due to clustering. In order to
completely account for the primary emission case for dijet
photoproduction one would need to generalise the calculations
presented here for a single $q\bar{q}$ dipole to the case of several
hard emitting dipoles. An exactly similar calculation would be
needed for the case of hadron-hadron collisions and this is work in
progress. It is straightforward however to at least estimate the
effect of our findings on the photoproduction case and we deal with
this in the following section.

\section{Gaps between jets at HERA -- the ZEUS analysis}

\FIGURE[ht]{\epsfig{file=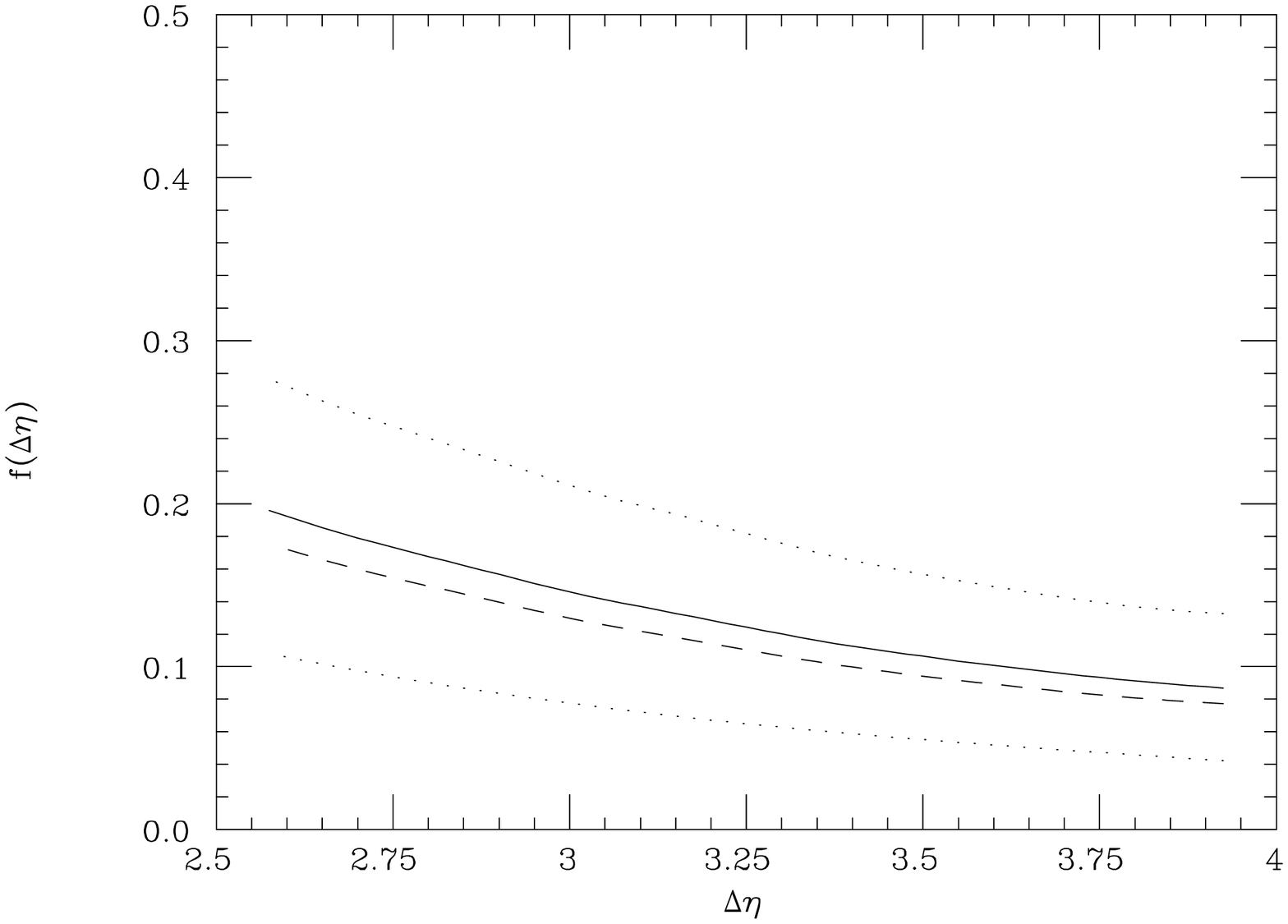,width =0.7 \textwidth, angle =90}
\caption{\label{Fig:ZEUS1}The gap fraction for the ZEUS analysis
with a $k_t$-defined final state  ($R=1.0$ and $Q_{\Omega}=0.5$
GeV). The solid line shows the effect of resummed primary emission,
the primary emission clustering correction factor and the non-global
suppression factor. The overall theoretical uncertainty in all three
contributions is shown by the dotted lines. The dashed line
indicates the gap fraction obtained by only including primary
resummed emission without accounting for clustering.}}
\FIGURE[ht]{\epsfig{file=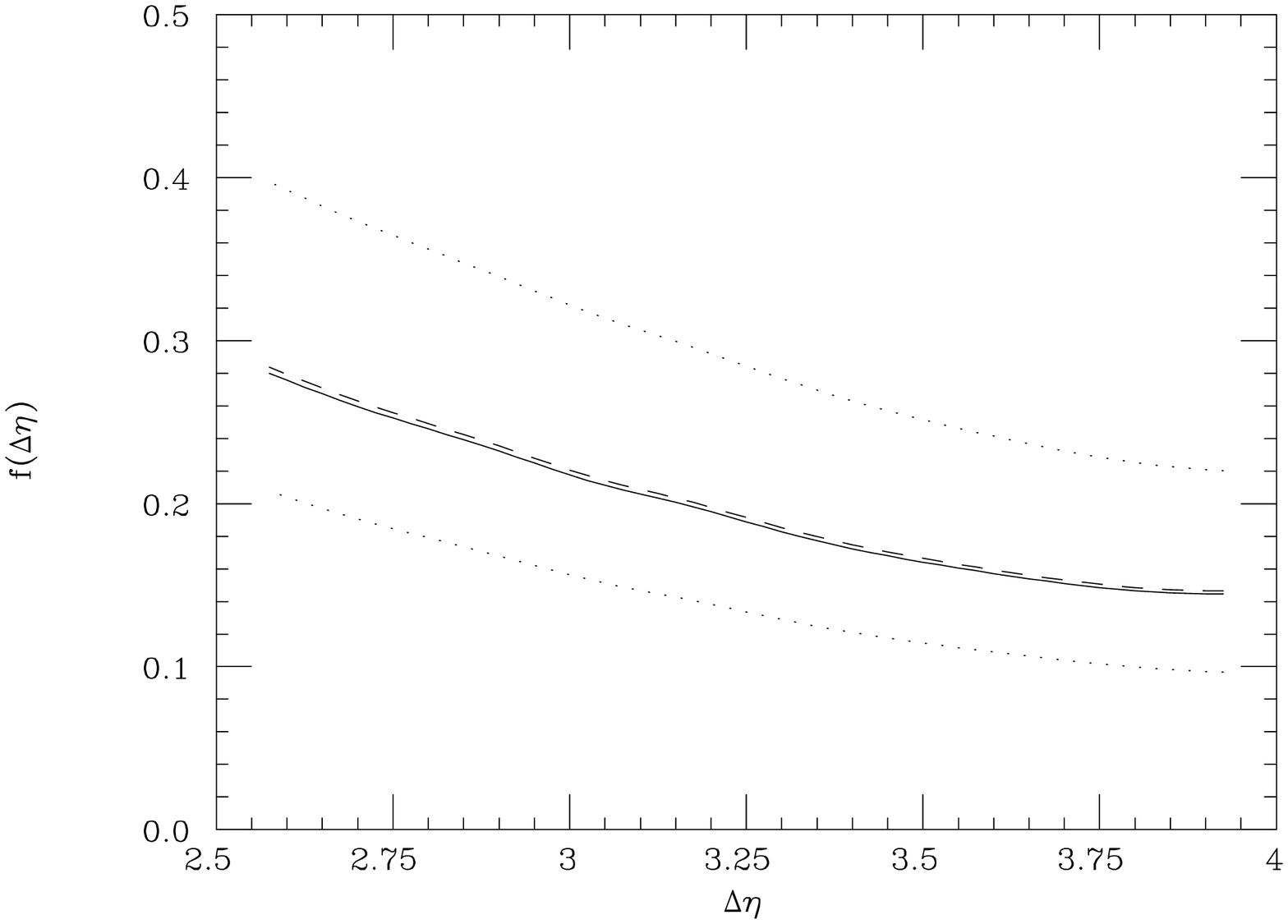,width =0.7 \textwidth, angle =90}
\caption{\label{Fig:ZEUS2}The gap fraction for the ZEUS analysis
with a kt-defined final state  ($R=1.0$ and $Q_{\Omega}=1.0$ GeV).
The solid line shows the effect of resummed primary emission, the
primary emission clustering correction factor and the non-global
suppression factor. The overall theoretical uncertainty in all three
contributions is shown by the dotted lines. The dashed line
indicates the gap fraction obtained by only including primary
resummed emission without accounting for clustering.}}

We can test the perturbative framework presented in this paper with
energy flow measurements in the photoproduction of dijets. These
energy flow observables are defined with two hard jets in the
central detector region separated by a gap in pseudorapidity. A gap
event is defined when the sum of the hadronic transverse energy in
the gap is less than a cut-off, and the gap fraction is defined as
the ratio of the gap cross-section to the total inclusive
cross-section. The energy flow observables measured by
ZEUS~\cite{ZEUS} and H1~\cite{h1} use the $k_t$ clustering
definition of the hadronic final state, and the transverse energy in
the gap is given by the sum of the mini-jet contributions. In this
paper we focus on the ZEUS measurements and provide revised
theoretical estimates for them. These revisions lead to changes that
are minor in the context of the overall theoretical uncertainty but
should become more significant once the matching to fixed
higher-orders is carried out and an estimate of the next-to--leading
logarithms is obtained. The H1 data was considered in
Ref.~\cite{AS2}, where the theoretical analysis consisted of only
the resummed primary emission contribution without taking account of
the effect of $k_t$ clustering.

The ZEUS data was obtained by colliding 27.5 GeV positrons with 820
GeV protons, with a total integrated luminosity of 38.6 $\pm$ 1.6
pb$^{-1}$ in the 1996-1997 HERA running period.  The full details of
the ZEUS analysis can be found in Ref.~\cite{ZEUS}, but the cuts
relevant to the calculations in this paper are:
\begin{eqnarray}
0.2 < y < 0.75\,, \nonumber \\
Q^2 < 1\,\mathrm{GeV}^2\,, \nonumber \\
6,5 \, \mathrm{GeV} < E_T(1,2)\,, \nonumber \\
|\eta(1,2)| < 2.4\,, \nonumber \\
|0.5(\eta_1 + \eta_2)| < 0.75\,, \nonumber \\
2.5 < \Delta\eta < 4\,,\nonumber
\end{eqnarray}
where $y$ is the inelasticity, $Q^2$ is the virtuality of the
photon, $E_T(1,2)$ are the transverse energies of the two hard jets,
$\eta(1,2)$ are the pseudorapidities of the two hardest jets and
$\Delta\eta$ is the jet rapidity difference. The further requirement
for the gap sample is $E_{t,\,\mathrm{gap}}\, <Q_\Omega=$ 0.5, 1,
1.5, 2 GeV, and the clustering parameter $R$ is always taken to be
unity.

The theoretical prediction for the gap fraction is composed of the
primary piece, with corrections due to clustering, and the
non-global piece. We shall now describe each in turn.

The resummed primary contribution ignoring the clustering
corrections, is obtained from the factorisation methods of Sterman
et al~\cite{KS} and is described in Ref.~\cite{AS2}. The four-jet
case of photoproduction requires a matrix formalism, and the
exponents of the Sudakov factors in the gap cross-section are
anomalous dimension matrices over the basis of possible colour flows
of the hard sub-process. The emission of soft gluons cause mixing of
the colour basis. Consideration of the eigenvectors and eigenvalues
of the anomalous dimension matrices, together with
sub-process--dependent hard and soft matrices, allows the resummed
four-jet primary emission differential cross-section to be written
as~\cite{AS2}:
\begin{equation}
\label{primary} \frac{d\sigma}{d\eta}=\sum_{L,I}H_{IL}S_{LI}\exp
\left\{\left(\lambda_L^{*} (\eta,\Omega)+\lambda_I
(\eta,\Omega)\right)\int_{p_t}^{Q_{\Omega}} \frac{d\mu}{\mu}
\alpha_s(\mu)\right\},
\end{equation}
where $H$ and $S$ denote the hard and soft matrices (expanded over
the colour basis), $\lambda$ denotes the eigenvalues of the
anomalous dimension matrices, $\eta=\Delta\eta/2$ and $p_t$ is the
hard scale of the process. This was computed in Ref.~\cite{AS2} for
the case of photoproduction and energy flow observables measured by
H1. In this paper we have recomputed this differential gap
cross-section for the observable defined by the ZEUS collaboration.
The uncertainty in the renormalisation scale is quantified by
varying the hard scale in the resummation by a factor of 2 (upper
bound) and 0.5 (lower bound).

We now need to account for the effect of clustering on
Eq.~\eqref{primary}. Since we do not have as yet the full results
for the four-jet case of photoproduction we simply estimate the full
correction as the square of the correction arising in the two-jet
case dealt with here, using the appropriate colour factors for each
hard sub-process. This was also the method used to approximate the
non-global contribution for the four-jet case in Ref.~\cite{AS2}.
While we emphasise that this is only a rough way of examining the
impact of the clustering dependent terms computed here, given the
size of the effects we are dealing with, it is clear that no
significant differences ought to emerge if one were to properly
compute the various dipole sub-processes we need to account for. We
also include the revised and virtually negligible non-global
component in an identical fashion to arrive at the best current
theoretical estimates.

The results for the ZEUS gap-fraction with a $k_t$-defined final
state are shown in Figs.~\ref{Fig:ZEUS1} and~\ref{Fig:ZEUS2}. We
consider here two different values for the gap energy $Q_\Omega$.
For the value of $Q_\Omega = 0.5$ GeV one notes that the full
prediction accounting approximately for all additional sources of
single-logarithmic enhancements, is somewhat higher than the pure
``Sudakov'' type prediction. This is due to the extra primary terms
we compute here, non-global corrections being negligible. For a
larger value of $Q_\Omega = 1.0$ GeV the difference between  the
clustered and unclustered primary results is negligible. We also
note the large theoretical uncertainty on the prediction as
represented by the renormalisation scale dependence. This is to be
expected in light of the fact that the predictions here are not
matched to fixed-order and account only for the leading logarithms.
Improvements along both these directions should be possible in the
immediate future after which the role of the various effects we
highlighted here should be revisited.

\section{Conclusions}

In the present paper we have shed further light on resummations of
$k_t$-clustered final states. We have shown that both the primary
and non-global components of the resummed result are affected by
clustering and dealt with the resummation of each in turn. For the
non-global component we find that the results after applying
clustering are different from those presented earlier~\cite{AS1}.
The new results we present here indicate an even smaller non-global
component than previously believed.

We have also shown how the primary emission clustering effects can
be resummed to all orders as an expansion in the clustering
parameter $R$ and computed a few terms of the series. The analytical
results we have provided here for a single emitting dipole should be
generalisable to the case of several hard dipoles (multi-jet
processes). This should then enable one to write a correct resummed
result for primary emissions to a high accuracy and deal with the
reduced non-global component in the large $N_c$ limit. Such progress
is relevant not just to energy-flow studies but to any jet
observable of a non-global nature, requiring resummation. An example
is the azimuthal angle $\Delta \phi_{jj}$ between jets, mentioned
previously. The work we have carried out should enable
next-to--leading log calculations of such jet observables to
sufficient accuracy to enable phenomenological studies of the same.

Lastly we have also mentioned the impact of the new findings on the
ZEUS gaps-between--jets analysis. Since the non-global effects are
very small for $R=1$ the main new effect is the additional
clustering dependent primary terms we computed here. Approximating
the effect of these terms for the case of photoproduction, somewhat
changes the theoretical predictions but this change is insignificant
given the large theoretical uncertainty that arises due to missing
higher orders and unaccounted for next-to--leading logarithms. We
consider both these areas as avenues for further work and hope that
more stringent comparisons can thus be made in the very near future.

\acknowledgments{ We would like to thank Gavin Salam and Claire
Gwenlan for helpful conversations. One of us (MD) would like to
thank the LPTHE Jussieu, Paris for their kind hospitality during the
completion of this work.}

\end{document}